\documentclass[%
prl,
reprint
]{revtex4-1}
\usepackage{graphicx}
\usepackage[update,prepend]{epstopdf}
\usepackage{bm}
\usepackage{hhline}
\usepackage{siunitx}
\usepackage{amsmath}

\DeclareGraphicsExtensions{.eps,.pdf}

\begin{document}

\title{Random telegraph signal analysis with a recurrent neural network}

\author{N. J. Lambert}
\email{nicholas.lambert@otago.ac.nz}
\affiliation{Department of Physics, University of Otago, Dunedin, New Zealand}
\affiliation{The Dodd-Walls Centre for Photonic and Quantum Technologies, New Zealand}
\author{A. A. Esmail}
\affiliation{Microelectronics Group, Cavendish Laboratory, University of Cambridge, Cambridge, CB3 0HE, UK}
\author{M. Edwards}
\affiliation{Microelectronics Group, Cavendish Laboratory, University of Cambridge, Cambridge, CB3 0HE, UK}
\author{A. J. Ferguson}
\affiliation{Microelectronics Group, Cavendish Laboratory, University of Cambridge, Cambridge, CB3 0HE, UK}
\author{H. G. L. Schwefel}
\affiliation{Department of Physics, University of Otago, Dunedin, New Zealand}
\affiliation{The Dodd-Walls Centre for Photonic and Quantum Technologies, New Zealand}

\date{\today}

\begin{abstract}
We use an artificial neural network to analyze asymmetric noisy random telegraph signals (RTSs), and extract underlying transition rates. We demonstrate that a long short-term memory neural network can vastly outperform conventional methods, particularly for noisy signals. Our technique gives reliable results as the signal-to-noise ratio approaches one, and over a wide range of underlying transition rates. We apply our method to random telegraph signals generated by a superconducting double dot based photon detector, allowing us to extend our measurement of quasiparticle dynamics to new temperature regimes.

\end{abstract}


\maketitle

An asymmetric random telegraph signal is a signal which stochastic switches between two levels $y=y_1$ and $y=y_2$. They are a common result of measurements on a wide variety of physical systems, including ion channels in cells \cite{hilleIonChannelsExcitable2001}, semiconductor devices such as transistors \cite{kandiahRandomTelegraphSignal1994,kandiahPhysicalModelRandom1989}, quantum dots \cite{efrosRandomTelegraphSignal1997} and optoelectronic devices \cite{wangRandomTelegraphSignal2006}, high-$T_c$ superconductors \cite{jungElementaryMacroscopicTwo1996}, and single-Cooper-pair boxes \cite{shawKineticsNonequilibriumQuasiparticle2008}, as well as being the building block of $1/f$ noise \cite{kirtonNoiseSolidstateMicrostructures1989}. The transition rates from $1$ ($2$) to $2$ ($1$), $\Gamma_{1(2)}$, are the accessible parameters describing the dynamics of the underlying system, and it is often desirable to extract them from the measured time sequence.

\begin{figure}
\includegraphics{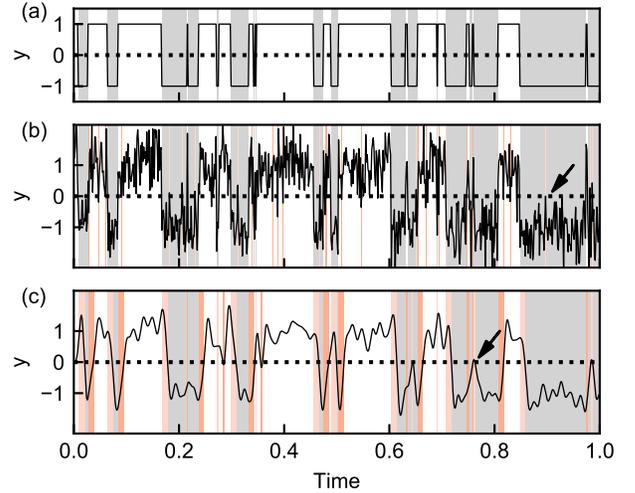}
\caption{Filtered noisy RTSs, with apparent periods in each state marked as grey and white. (a) The underlying signal has periods in states 1 and 2, corresponding to signal levels $\pm1$. (b) The same signal with added Gaussian white noise of standard deviation 0.4. The additional noise creates false excursions from one state into the other (marked in red, example arrowed). (c) The signal in (b) passed through a digital low pass filter. Now short periods in one state can be missed (marked in red, example arrowed).}\label{fig:problems}

\end{figure}

The most straightforward way to do this is to sample the time domain signal at some rate $f_s$, divide it into periods in each of states $1$ and $2$ (Fig.~\ref{fig:problems}(a)), histogram the dwell times $\tau_{1(2)}$ and fit $k e^{-\Gamma_{1(2)}\tau_{1(2)}}$ to the resulting distribution. However, the presence of noise and a finite measurement bandwidth will result in the measured statistics not representing the underlying system accurately. The problem is two-fold: noise whilst in one state can result in a \emph{false} time period in the other state being detected (Fig.~\ref{fig:problems}(b)), and a limited bandwidth means that \emph{genuine} short-period excursions to the other state are not seen (Fig.~\ref{fig:problems}(c)). This later effect also joins together the two time periods either side of the missed period, resulting in a \emph{false} long period. 

A variety of solutions to this problem have been proposed. Some focus on optimising the threshold at which the signal is divided into states 1 and 2 \cite{yuzhelevskiRandomTelegraphNoise2000}. Naaman and Aumentado modelled the detector as a separate process \cite{naamanPoissonTransitionRates2006}, and derived corrections to the measured rates. Other techniques include wavelet edge detection \cite{pranceIdentifyingSingleElectron2015}, autocorrelation methods \cite{martin-martinezNewWeightedTime2014}, cross corelation methods \cite{kungTimeresolvedChargeDetection2009}, and analysis of the probability density function of the signal \cite{lambertMicrowaveIrradiationQuasiparticles2017, singhDeterminingParametersRandom2018}.

In this Letter, we demonstrate that a recurrent neural network can be used to extract the underlying rates from noisy, bandwidth limited random telegraph signals. Neural networks (NNs) comprise an input layer holding the data to be analyzed, an output layer giving the result of the analysis, and one or more `hidden' layers of any number of nodes. The connections between nodes can be arranged in a wide variety of different topologies, depending on the nature of the analysis to be performed. The weights of the internode connections are tuned using gradient descent methods to achieve the optimal mapping between the possible input vectors and the desired output vectors, a process referred to as `training'. Neural networks have proven to be a versatile tool, capable of tackling diverse problems including image recognition, financial fraud detection and natural language processing. They have also proven useful for the physical sciences, and have been applied to astrophysical images \cite{banerjiGalaxyZooReproducing2010,hezavehFastAutomatedAnalysis2017}, meta-analysis of the scientific literature \cite{tshitoyanUnsupervisedWordEmbeddings2019}, the generation of quantum error correction algorithms \cite{foselReinforcementLearningNeural2018}, and the characterization of quantum dots \cite{kalantreMachineLearningTechniques2019} and dopants \cite{usmanAtomiclevelCharacterisationQuantum2019} in semiconductors.

Recurrent NNs are a class of NN which act sequentially in a particular direction along an input data array, with an internal memory allowing correlations between data points to affect the output. They are therefore particularly suitable for time sequence data. We use a long short-term memory (LSTM) \cite{hochreiterLongShorttermMemory1997} architecture, a recurrent NN designed for time sequences in which related information can have significant temporal separation, such as speech \cite{gravesSpeechRecognitionDeep2013} and handwriting \cite{gravesUnconstrainedOnlineHandwriting2008} recognition, and musical analysis \cite{eckLearningLongTermStructure2002}.

\begin{table}[]
\caption{\label{tab:network}The configuration of our neural network. }

\begin{tabular}{cccc}
\hhline{====}
Layer  & Type            & Size   & Activation \\ \hline
Input  & -               & $10^5$ & -          \\
1      & LSTM            & 128    & sigmoid    \\
2      & Fully connected & 128    & elu        \\
3      & Fully connected & 128    & elu        \\
4      & Fully connected & 128    & elu        \\
5      & Fully connected & 10     & elu        \\
Output & -               & 1      & relu       \\ \hhline{====}
\end{tabular}
\end{table}

Our NN has an input size of $10^5$ time samples, followed by an LSTM layer with internal size 128 (Table~\ref{tab:network}). The subsequent fully connected hidden layers are of sizes 128, 128, 128 and 10, with elu activation \cite{clevertFastAccurateDeep2015} functions. The output uses relu activation to ensure a non-negative output value, and is single valued; we exploit the symmetry of the RTS by training for only one rate, with the other accessible by simply inverting the signal, such that $y(t) \rightarrow -y(t)$. This gives faster and more accurate training.
	
We implement the NN in Python 3.7 using Keras~\cite{chollet2015keras}, with the Tensorflow back end. Training and prediction is accelerated by the use of a Nvidia Tesla K40c GPU. The NN was typically trained over 250 epochs, with 100 steps per epoch, taking around 13.6 hours. The NN is trained using synthetic RTSs, generated with independent rates $\Gamma_1$ and $\Gamma_2$ uniformly distributed on the interval between $10^{-3}\cdot f_s$ and $f_s$. Rather than $\Gamma_{1,2}$, we train for $\log_{10}(\Gamma_{1,2})$, to compress the output space.

We find that a realistic noise model is necessary for accurate analysis of real data. Two components of additive noise are generated: one with a $1/f$ power spectrum and randomized phase, representing, for example, noise processes in a semiconductor substrate; and one with a flat power spectrum and randomized phase, representing instrumentation noise. The amplitude of the noise added to the training data can be fixed to reflect the measured experimental noise, or varied over a wider distribution. We find that, in general, the NN cannot be successfully trained if presented with noisy data initially. Instead the training is started with noise free data, and the noise amplitude (or range of amplitudes) is increased every 20 training epochs until the desired level is reached. Finally, the generated signals are normalized such that they have mean 0 and standard deviation 1.

\begin{figure*}
\includegraphics{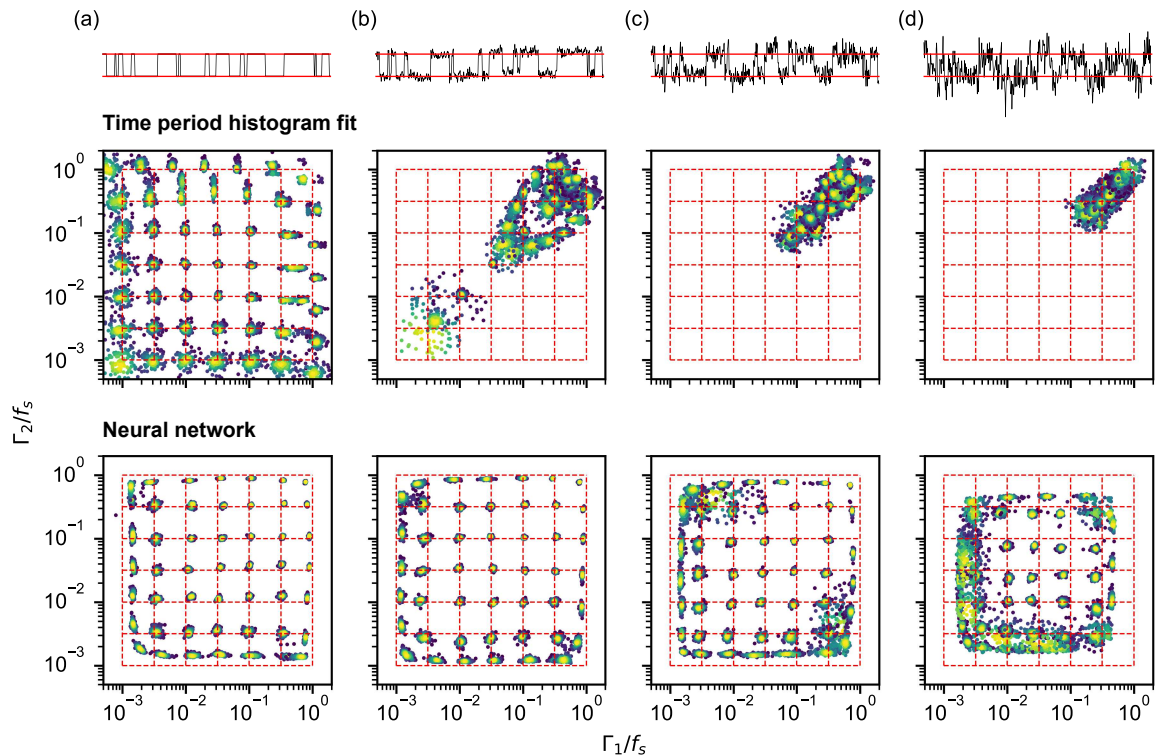}
\caption{Application of the trained neural network to synthetic data with varying additive noise. Top panels, example RTSs with levels $\pm 1$ (red lines) and noise of standard deviations (a) 0, (b) 0.2, (c) 0.4, and (d) 0.6. Middle panels, analysis of RTSs with rates $10^{-3} f_s\leq\Gamma_{1,2}\leq f_s$ using threshold binning of time intervals, and a fit to an exponential decay. The red grid marks the underlying rates of the synthesised data. The addition of only a small amount of noise causes the method to fail completely. Bottom panels show analysis of the same RTSs using the trained neural network, showing greatly increased robustness against noise.}\label{fig:synthresults}

\end{figure*}

Once trained, the effectiveness of the NN can be tested by applying it to RTS data sequences. In Fig.~\ref{fig:synthresults} we compare the results of testing our NN on synthetic data of length $10^5$ samples to alternative methods. The NN is applied to RTSs with logarithmically spaced transition rates in the range $10^{-3}f_s <\Gamma_{1,2}<f_s$, signal levels $y=\pm 1$ and noise with standard deviations 0 (no additive noise), 0.2, 0.4, and 0.6 (examples in upper panels). In Fig.~\ref{fig:synthresults} a-d we show the results of analysis of 100 RTSs for each rate pair $(\Gamma_1 ,\Gamma_2)$ by the time period histogram method (middle panels) and our NN (lower panels). The red dashed grid marks the underlying rates of the synthesised signals.

The NN is more effective for all noise levels. It is less accurate when one or both rates are low, and so there are fewer events to analyze, or equivalently the total power in the signal is low. Signals with higher rates and higher noises are also challenging, reflecting the similarity between a short period in a particular state and a spike due to noise. Nevertheless, the NN still performs well for regimes in which the time period histogram method fails completely, being particularly vulnerable to errors caused by $1/f$ noise.

\begin{figure*}
\includegraphics{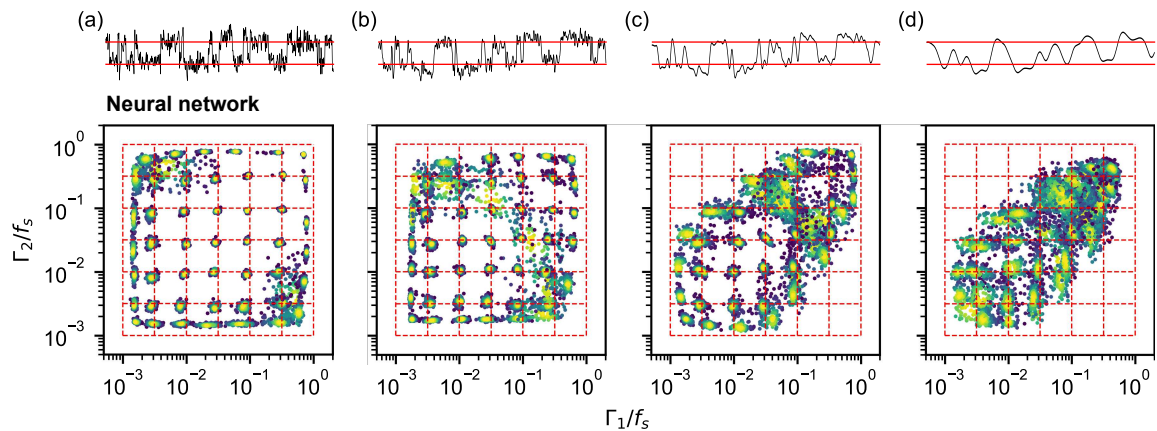}
\caption{Application of the trained neural network to synthetic data with noise standard deviation 0.4, passed through digital low-pass filters with cutoff frequencies (a) $f_c > f_s$ (identical to Fig.~\ref{fig:synthresults}(c)) (b) $f_c=f_s/3.162$ (c) $f_c=f_s/10$ and (d) $f_c=f_s/31.62$. Example time traces are shown in the top panels, and extracted rates in the lower panels.}\label{fig:synthresultscutoff}

\end{figure*}

In Fig.~\ref{fig:synthresultscutoff} we study the effect of filtering the RTS. The noise level for synthetic signals is now fixed at 0.4, and they are filtered using a 5th order digital Butterworth filter with \SI{3}{\decibel} cutoffs of $f_c=f_s/3.162$, $f_c=f_s/10$ and  $f_c=f_s/31.62$, and again analyzed using appropriately trained NNs for each cutoff frequency. The NN is generally robust against low pass filtering when $f_c > \Gamma_1 + \Gamma_2$. If the rates are outside this regime, a significant spectral content of the signal is above the passband of the filter.\\

\begin{figure}
\includegraphics{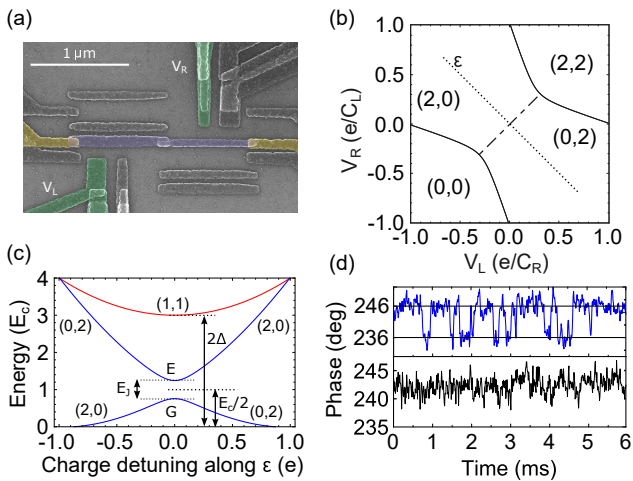}
\caption{(a) False color SEM of a superconducting double dot. Purple regions are the superconducting islands, yellow regions are normal-metal grounded reservoirs, and electrostatic control gates are green. Other gates are not used in these experiments. (b) The charge stability diagram, showing the ground state charge configuration as a function of gate voltages $V_L$ and $V_R$. (c) Charge state energies along $\epsilon$ in panel (b). The anticrossing between the (2,0) and (0,2) charge states is mediated by the Josephson energy of the interdot tunnel junction, while the energy of the (1,1) state is due to the $\Delta$. (d) Top panel - r.f.~reflected phase at B = 0, T = \SI{25}{\milli\kelvin}, $\epsilon = 0$, showing an RTS. Bottom panel - r.f.~reflected phase at B = \SI{180}{\milli\tesla}, T = \SI{100}{\milli\kelvin}, $\epsilon = 0$ with no RTS apparent.}\label{fig:SCDD}

\end{figure}

	
We now apply our neural network to RTSs due to microwave photon absorption processes in a superconducting double dot (SCDD) \cite{lambertExperimentalObservationBreaking2014, lambertMicrowaveIrradiationQuasiparticles2017a}. The device comprises two aluminium superconducting islands coupled to each other by a Josephson junction, and to metallic leads by Superconducting-Insulator-Normal tunnel junctions (Fig.~\ref{fig:SCDD}(a)). The charge state of the device is described by the differences in charge on each island from some arbitrarily chosen even charge state, $q_\textrm{left}$ and $q_\textrm{right}$, and is labelled ($q_\textrm{left}/e$, $q_\textrm{right}/e$). The behaviour of the device is governed by competition between the superconducting gap $\Delta$ and the charging energy $E_c$. When $\Delta>3E_c/4$ the device is protected from quasiparticle excitations and the charge residing on each island is quantised in units of $2e$.

The energy curvature associated with the anticrossing between the (0,2) and (2,0) charge states gives the device a finite quantum capacitance, but this can be removed by the destruction of coherence due to the presence of unpaired electrons, a process known as quasiparticle poisoning~\cite{naamanTimedomainMeasurementsQuasiparticle2006,fergusonMicrosecondResolutionQuasiparticle2006,lutchynQuasiparticleDecayRate2005}. By monitoring the quantum capacitance of the SCDD via radiofrequency reflectrometry, Cooper pair breaking and reforming events can be observed. Time domain measurements (Fig.~\ref{fig:SCDD}(d), upper panel) yield an RTS with one rate determined by the vulnerability of the Cooper pairs to incident photons, and the other determined by the recombination dynamics.

Because the transitions between charge configurations are driven by the energy difference between the states, it is interesting to study the transition rates as a function of temperature and applied magnetic field, which affect the state's free energy via changes in entropy and the superconducting gap. For low temperatures ($T=\SI{35}{\milli\kelvin}$) and magnetic fields ($B<\SI{120}{\milli\tesla}$), rates can be determined by applying thresholding methods to the measured signal. But at higher temperatures and fields, the quality of the phase signal is sufficiently degraded (Fig.~\ref{fig:SCDD}(d), lower panel) that this technique does not work.

\begin{figure}
\includegraphics{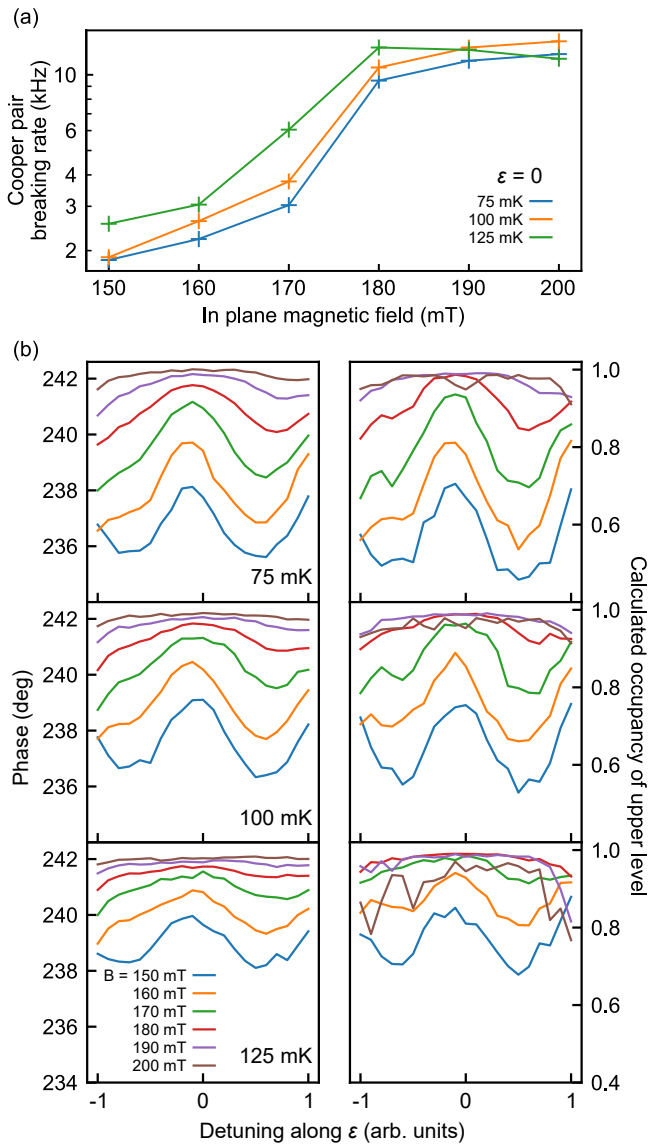}
\caption{Application of trained NN to experimental data from an SCDD. (a) Cooper pair breaking rates as a function of applied field for increasing temperatures at zero detuning from the charge state anticrossing. (b) Left panels - time average phase as a function of $\epsilon$ detuning. The measured phase is a measure of the proportion of time spent in the excited state. Right panels - the occupancy of the excited state deduced from the rates $\Gamma_1$ and $\Gamma_2$ extracted by the trained NN.}\label{fig:realresults}

\end{figure}

Time traces  with an acquisition time of \SI{1}{\second} and comprising $10^5$ phase measurements were taken at temperatures of \SI{75}{\milli\kelvin}, \SI{100}{\milli\kelvin} and \SI{125}{\milli\kelvin}, and at fields of $\SI{150}{\milli\tesla}<H<\SI{200}{\milli\tesla}$. The demodulated phase signal was filtered with a cutoff of \SI{15}{\kilo\hertz} before sampling. For each field and temperature, measurements were made at points along a line in the charge stability diagram corresponding to transfer of charge from one island to the other, labelled $\epsilon$ in Fig.~\ref{fig:SCDD}(b). Traces were found to have a low signal-to-noise ratio, with no obvious random telegraph behaviour (Fig.~\ref{fig:SCDD}(d), lower panel). Rates can nevertheless be extracted using our NN. In Fig.~\ref{fig:realresults} we plot the rates at zero detuning for increasing field at three temperatures, and observe the super-exponential behaviour previously seen at lower temperatures 

To validate the rates extracted using our NN, we compare the mean phase for each measured trace with the expected mean value for an RTS having extracted rates $\Gamma_1$ and $\Gamma_2$, 
\begin{align}
\bar{y} = \frac{\Gamma_1/\Gamma_2}{\Gamma_1/\Gamma_2+1}.
\end{align}
In each case this is a measure of the excited state occupancy \cite{esmailCooperPairTunnelling2017}. In Fig.~\ref{fig:realresults}(b) we plot the measured mean signals (left) and deduced excited state occupancy (right). The agreement is excellent, demonstrating the efficacy of our NN for analysis of experimental data.

In summary, we find that an LSTM recurrent neural network is a powerful tool for the determination of the transition rates underlying noisy random telegraph signals with finite sampling rates. The network architecture is simple, yet versatile enough to apply to different signal parameters, and can be trained in a reasonable time on relatively modest hardware. This has allowed us to analyze previously inaccessible signals, and is particularly useful for measurement of delicate quantum systems for which measurements are difficult, and the SNR likely to be low. \\

We would like to acknowledge support from the MBIE (NZ) Endeavour Smart Ideas fund, Hitachi Cambridge Laboratory, and UK EPSRC Grant No. EP/K027018/1. A.J.F. was supported by a Hitachi Research fellowship.

\bibliography{RTSNN}

\begin{thebibliography}{33}%
\makeatletter
\providecommand \@ifxundefined [1]{%
 \@ifx{#1\undefined}
}%
\providecommand \@ifnum [1]{%
 \ifnum #1\expandafter \@firstoftwo
 \else \expandafter \@secondoftwo
 \fi
}%
\providecommand \@ifx [1]{%
 \ifx #1\expandafter \@firstoftwo
 \else \expandafter \@secondoftwo
 \fi
}%
\providecommand \natexlab [1]{#1}%
\providecommand \enquote  [1]{``#1''}%
\providecommand \bibnamefont  [1]{#1}%
\providecommand \bibfnamefont [1]{#1}%
\providecommand \citenamefont [1]{#1}%
\providecommand \href@noop [0]{\@secondoftwo}%
\providecommand \href [0]{\begingroup \@sanitize@url \@href}%
\providecommand \@href[1]{\@@startlink{#1}\@@href}%
\providecommand \@@href[1]{\endgroup#1\@@endlink}%
\providecommand \@sanitize@url [0]{\catcode `\\12\catcode `\$12\catcode
  `\&12\catcode `\#12\catcode `\^12\catcode `\_12\catcode `\%12\relax}%
\providecommand \@@startlink[1]{}%
\providecommand \@@endlink[0]{}%
\providecommand \url  [0]{\begingroup\@sanitize@url \@url }%
\providecommand \@url [1]{\endgroup\@href {#1}{\urlprefix }}%
\providecommand \urlprefix  [0]{URL }%
\providecommand \Eprint [0]{\href }%
\providecommand \doibase [0]{http://dx.doi.org/}%
\providecommand \selectlanguage [0]{\@gobble}%
\providecommand \bibinfo  [0]{\@secondoftwo}%
\providecommand \bibfield  [0]{\@secondoftwo}%
\providecommand \translation [1]{[#1]}%
\providecommand \BibitemOpen [0]{}%
\providecommand \bibitemStop [0]{}%
\providecommand \bibitemNoStop [0]{.\EOS\space}%
\providecommand \EOS [0]{\spacefactor3000\relax}%
\providecommand \BibitemShut  [1]{\csname bibitem#1\endcsname}%
\let\auto@bib@innerbib\@empty
\bibitem [{\citenamefont {Hille}(2001)}]{hilleIonChannelsExcitable2001}%
  \BibitemOpen
  \bibfield  {author} {\bibinfo {author} {\bibfnamefont {B.}~\bibnamefont
  {Hille}},\ }\href@noop {} {\emph {\bibinfo {title} {Ion {{Channels}} of
  {{Excitable Membranes}}}}},\ \bibinfo {edition} {third edition}\ ed.\
  (\bibinfo  {publisher} {{Sinauer Associates}},\ \bibinfo {address}
  {{Sunderland, Massachusetts}},\ \bibinfo {year} {2001})\BibitemShut {NoStop}%
\bibitem [{\citenamefont {Kandiah}(1994)}]{kandiahRandomTelegraphSignal1994}%
  \BibitemOpen
  \bibfield  {author} {\bibinfo {author} {\bibfnamefont {K.}~\bibnamefont
  {Kandiah}},\ }\href {\doibase 10.1109/16.333818} {\bibfield  {journal}
  {\bibinfo  {journal} {IEEE Transactions on Electron Devices}\ }\textbf
  {\bibinfo {volume} {41}},\ \bibinfo {pages} {2006} (\bibinfo {year}
  {1994})}\BibitemShut {NoStop}%
\bibitem [{\citenamefont {Kandiah}\ \emph {et~al.}(1989)\citenamefont
  {Kandiah}, \citenamefont {Deighton},\ and\ \citenamefont
  {Whiting}}]{kandiahPhysicalModelRandom1989}%
  \BibitemOpen
  \bibfield  {author} {\bibinfo {author} {\bibfnamefont {K.}~\bibnamefont
  {Kandiah}}, \bibinfo {author} {\bibfnamefont {M.~O.}\ \bibnamefont
  {Deighton}}, \ and\ \bibinfo {author} {\bibfnamefont {F.~B.}\ \bibnamefont
  {Whiting}},\ }\href@noop {} {\bibfield  {journal} {\bibinfo  {journal}
  {Journal of Applied Physics}\ }\textbf {\bibinfo {volume} {66}},\ \bibinfo
  {pages} {937} (\bibinfo {year} {1989})}\BibitemShut {NoStop}%
\bibitem [{\citenamefont {Efros}\ and\ \citenamefont
  {Rosen}(1997)}]{efrosRandomTelegraphSignal1997}%
  \BibitemOpen
  \bibfield  {author} {\bibinfo {author} {\bibfnamefont {A.~L.}\ \bibnamefont
  {Efros}}\ and\ \bibinfo {author} {\bibfnamefont {M.}~\bibnamefont {Rosen}},\
  }\href {\doibase 10.1103/PhysRevLett.78.1110} {\bibfield  {journal} {\bibinfo
   {journal} {Physical Review Letters}\ }\textbf {\bibinfo {volume} {78}},\
  \bibinfo {pages} {1110} (\bibinfo {year} {1997})}\BibitemShut {NoStop}%
\bibitem [{\citenamefont {Wang}\ \emph {et~al.}(2006)\citenamefont {Wang},
  \citenamefont {Rao}, \citenamefont {Mierop},\ and\ \citenamefont
  {Theuwissen}}]{wangRandomTelegraphSignal2006}%
  \BibitemOpen
  \bibfield  {author} {\bibinfo {author} {\bibfnamefont {X.}~\bibnamefont
  {Wang}}, \bibinfo {author} {\bibfnamefont {P.~R.}\ \bibnamefont {Rao}},
  \bibinfo {author} {\bibfnamefont {A.}~\bibnamefont {Mierop}}, \ and\ \bibinfo
  {author} {\bibfnamefont {A.~J.}\ \bibnamefont {Theuwissen}},\ }in\ \href
  {\doibase 10.1109/IEDM.2006.346973} {\emph {\bibinfo {booktitle} {2006
  {{International Electron Devices Meeting}}}}}\ (\bibinfo {year} {2006})\ pp.\
  \bibinfo {pages} {1--4}\BibitemShut {NoStop}%
\bibitem [{\citenamefont {Jung}\ and\ \citenamefont
  {Savo}(1996)}]{jungElementaryMacroscopicTwo1996}%
  \BibitemOpen
  \bibfield  {author} {\bibinfo {author} {\bibfnamefont {G.}~\bibnamefont
  {Jung}}\ and\ \bibinfo {author} {\bibfnamefont {B.}~\bibnamefont {Savo}},\
  }\href {\doibase 10.1063/1.363150} {\bibfield  {journal} {\bibinfo  {journal}
  {Journal of Applied Physics}\ }\textbf {\bibinfo {volume} {80}},\ \bibinfo
  {pages} {2939} (\bibinfo {year} {1996})}\BibitemShut {NoStop}%
\bibitem [{\citenamefont {Shaw}\ \emph {et~al.}(2008)\citenamefont {Shaw},
  \citenamefont {Lutchyn}, \citenamefont {Delsing},\ and\ \citenamefont
  {Echternach}}]{shawKineticsNonequilibriumQuasiparticle2008}%
  \BibitemOpen
  \bibfield  {author} {\bibinfo {author} {\bibfnamefont {M.~D.}\ \bibnamefont
  {Shaw}}, \bibinfo {author} {\bibfnamefont {R.~M.}\ \bibnamefont {Lutchyn}},
  \bibinfo {author} {\bibfnamefont {P.}~\bibnamefont {Delsing}}, \ and\
  \bibinfo {author} {\bibfnamefont {P.~M.}\ \bibnamefont {Echternach}},\ }\href
  {\doibase 10.1103/PhysRevB.78.024503} {\bibfield  {journal} {\bibinfo
  {journal} {Physical Review B}\ }\textbf {\bibinfo {volume} {78}},\ \bibinfo
  {pages} {024503} (\bibinfo {year} {2008})}\BibitemShut {NoStop}%
\bibitem [{\citenamefont {Kirton}\ and\ \citenamefont
  {Uren}(1989)}]{kirtonNoiseSolidstateMicrostructures1989}%
  \BibitemOpen
  \bibfield  {author} {\bibinfo {author} {\bibfnamefont {M.~J.}\ \bibnamefont
  {Kirton}}\ and\ \bibinfo {author} {\bibfnamefont {M.~J.}\ \bibnamefont
  {Uren}},\ }\href {\doibase 10.1080/00018738900101122} {\bibfield  {journal}
  {\bibinfo  {journal} {Advances in Physics}\ }\textbf {\bibinfo {volume}
  {38}},\ \bibinfo {pages} {367} (\bibinfo {year} {1989})}\BibitemShut
  {NoStop}%
\bibitem [{\citenamefont {Yuzhelevski}\ \emph {et~al.}(2000)\citenamefont
  {Yuzhelevski}, \citenamefont {Yuzhelevski},\ and\ \citenamefont
  {Jung}}]{yuzhelevskiRandomTelegraphNoise2000}%
  \BibitemOpen
  \bibfield  {author} {\bibinfo {author} {\bibfnamefont {Y.}~\bibnamefont
  {Yuzhelevski}}, \bibinfo {author} {\bibfnamefont {M.}~\bibnamefont
  {Yuzhelevski}}, \ and\ \bibinfo {author} {\bibfnamefont {G.}~\bibnamefont
  {Jung}},\ }\href {\doibase 10.1063/1.1150519} {\bibfield  {journal} {\bibinfo
   {journal} {Review of Scientific Instruments}\ }\textbf {\bibinfo {volume}
  {71}},\ \bibinfo {pages} {1681} (\bibinfo {year} {2000})}\BibitemShut
  {NoStop}%
\bibitem [{\citenamefont {Naaman}\ and\ \citenamefont
  {Aumentado}(2006{\natexlab{a}})}]{naamanPoissonTransitionRates2006}%
  \BibitemOpen
  \bibfield  {author} {\bibinfo {author} {\bibfnamefont {O.}~\bibnamefont
  {Naaman}}\ and\ \bibinfo {author} {\bibfnamefont {J.}~\bibnamefont
  {Aumentado}},\ }\href {\doibase 10.1103/PhysRevLett.96.100201} {\bibfield
  {journal} {\bibinfo  {journal} {Physical Review Letters}\ }\textbf {\bibinfo
  {volume} {96}},\ \bibinfo {pages} {100201} (\bibinfo {year}
  {2006}{\natexlab{a}})}\BibitemShut {NoStop}%
\bibitem [{\citenamefont {Prance}\ \emph {et~al.}(2015)\citenamefont {Prance},
  \citenamefont {Bael}, \citenamefont {Simmons}, \citenamefont {Savage},
  \citenamefont {Lagally}, \citenamefont {Friesen}, \citenamefont
  {Coppersmith},\ and\ \citenamefont
  {Eriksson}}]{pranceIdentifyingSingleElectron2015}%
  \BibitemOpen
  \bibfield  {author} {\bibinfo {author} {\bibfnamefont {J.~R.}\ \bibnamefont
  {Prance}}, \bibinfo {author} {\bibfnamefont {B.~J.~V.}\ \bibnamefont {Bael}},
  \bibinfo {author} {\bibfnamefont {C.~B.}\ \bibnamefont {Simmons}}, \bibinfo
  {author} {\bibfnamefont {D.~E.}\ \bibnamefont {Savage}}, \bibinfo {author}
  {\bibfnamefont {M.~G.}\ \bibnamefont {Lagally}}, \bibinfo {author}
  {\bibfnamefont {M.}~\bibnamefont {Friesen}}, \bibinfo {author} {\bibfnamefont
  {S.~N.}\ \bibnamefont {Coppersmith}}, \ and\ \bibinfo {author} {\bibfnamefont
  {M.~A.}\ \bibnamefont {Eriksson}},\ }\href {\doibase
  10.1088/0957-4484/26/21/215201} {\bibfield  {journal} {\bibinfo  {journal}
  {Nanotechnology}\ }\textbf {\bibinfo {volume} {26}},\ \bibinfo {pages}
  {215201} (\bibinfo {year} {2015})}\BibitemShut {NoStop}%
\bibitem [{\citenamefont {{Martin-Martinez}}\ \emph {et~al.}(2014)\citenamefont
  {{Martin-Martinez}}, \citenamefont {Diaz}, \citenamefont {Rodriguez},
  \citenamefont {Nafria},\ and\ \citenamefont
  {Aymerich}}]{martin-martinezNewWeightedTime2014}%
  \BibitemOpen
  \bibfield  {author} {\bibinfo {author} {\bibfnamefont {J.}~\bibnamefont
  {{Martin-Martinez}}}, \bibinfo {author} {\bibfnamefont {J.}~\bibnamefont
  {Diaz}}, \bibinfo {author} {\bibfnamefont {R.}~\bibnamefont {Rodriguez}},
  \bibinfo {author} {\bibfnamefont {M.}~\bibnamefont {Nafria}}, \ and\ \bibinfo
  {author} {\bibfnamefont {X.}~\bibnamefont {Aymerich}},\ }\href {\doibase
  10.1109/LED.2014.2304673} {\bibfield  {journal} {\bibinfo  {journal} {IEEE
  Electron Device Letters}\ }\textbf {\bibinfo {volume} {35}},\ \bibinfo
  {pages} {479} (\bibinfo {year} {2014})}\BibitemShut {NoStop}%
\bibitem [{\citenamefont {K{\"u}ng}\ \emph {et~al.}(2009)\citenamefont
  {K{\"u}ng}, \citenamefont {Pf{\"a}ffli}, \citenamefont {Gustavsson},
  \citenamefont {Ihn}, \citenamefont {Ensslin}, \citenamefont {Reinwald},\ and\
  \citenamefont {Wegscheider}}]{kungTimeresolvedChargeDetection2009}%
  \BibitemOpen
  \bibfield  {author} {\bibinfo {author} {\bibfnamefont {B.}~\bibnamefont
  {K{\"u}ng}}, \bibinfo {author} {\bibfnamefont {O.}~\bibnamefont
  {Pf{\"a}ffli}}, \bibinfo {author} {\bibfnamefont {S.}~\bibnamefont
  {Gustavsson}}, \bibinfo {author} {\bibfnamefont {T.}~\bibnamefont {Ihn}},
  \bibinfo {author} {\bibfnamefont {K.}~\bibnamefont {Ensslin}}, \bibinfo
  {author} {\bibfnamefont {M.}~\bibnamefont {Reinwald}}, \ and\ \bibinfo
  {author} {\bibfnamefont {W.}~\bibnamefont {Wegscheider}},\ }\href {\doibase
  10.1103/PhysRevB.79.035314} {\bibfield  {journal} {\bibinfo  {journal}
  {Physical Review B}\ }\textbf {\bibinfo {volume} {79}},\ \bibinfo {pages}
  {035314} (\bibinfo {year} {2009})}\BibitemShut {NoStop}%
\bibitem [{\citenamefont {Lambert}\ \emph
  {et~al.}(2017{\natexlab{a}})\citenamefont {Lambert}, \citenamefont {Esmail},
  \citenamefont {Pollock}, \citenamefont {Edwards}, \citenamefont {Lovett},\
  and\ \citenamefont
  {Ferguson}}]{lambertMicrowaveIrradiationQuasiparticles2017}%
  \BibitemOpen
  \bibfield  {author} {\bibinfo {author} {\bibfnamefont {N.~J.}\ \bibnamefont
  {Lambert}}, \bibinfo {author} {\bibfnamefont {A.~A.}\ \bibnamefont {Esmail}},
  \bibinfo {author} {\bibfnamefont {F.~A.}\ \bibnamefont {Pollock}}, \bibinfo
  {author} {\bibfnamefont {M.}~\bibnamefont {Edwards}}, \bibinfo {author}
  {\bibfnamefont {B.~W.}\ \bibnamefont {Lovett}}, \ and\ \bibinfo {author}
  {\bibfnamefont {A.~J.}\ \bibnamefont {Ferguson}},\ }\href {\doibase
  10.1103/PhysRevB.95.235413} {\bibfield  {journal} {\bibinfo  {journal}
  {Physical Review B}\ }\textbf {\bibinfo {volume} {95}},\ \bibinfo {pages}
  {235413} (\bibinfo {year} {2017}{\natexlab{a}})}\BibitemShut {NoStop}%
\bibitem [{\citenamefont {Singh}\ \emph {et~al.}(2018)\citenamefont {Singh},
  \citenamefont {Mannila}, \citenamefont {Golubev}, \citenamefont {Peltonen},\
  and\ \citenamefont {Pekola}}]{singhDeterminingParametersRandom2018}%
  \BibitemOpen
  \bibfield  {author} {\bibinfo {author} {\bibfnamefont {S.}~\bibnamefont
  {Singh}}, \bibinfo {author} {\bibfnamefont {E.~T.}\ \bibnamefont {Mannila}},
  \bibinfo {author} {\bibfnamefont {D.~S.}\ \bibnamefont {Golubev}}, \bibinfo
  {author} {\bibfnamefont {J.~T.}\ \bibnamefont {Peltonen}}, \ and\ \bibinfo
  {author} {\bibfnamefont {J.~P.}\ \bibnamefont {Pekola}},\ }\href {\doibase
  10.1063/1.5033560} {\bibfield  {journal} {\bibinfo  {journal} {Applied
  Physics Letters}\ }\textbf {\bibinfo {volume} {112}},\ \bibinfo {pages}
  {243101} (\bibinfo {year} {2018})}\BibitemShut {NoStop}%
\bibitem [{\citenamefont {Banerji}\ \emph {et~al.}(2010)\citenamefont
  {Banerji}, \citenamefont {Lahav}, \citenamefont {Lintott}, \citenamefont
  {Abdalla}, \citenamefont {Schawinski}, \citenamefont {Bamford}, \citenamefont
  {Andreescu}, \citenamefont {Murray}, \citenamefont {Raddick}, \citenamefont
  {Slosar}, \citenamefont {Szalay}, \citenamefont {Thomas},\ and\ \citenamefont
  {Vandenberg}}]{banerjiGalaxyZooReproducing2010}%
  \BibitemOpen
  \bibfield  {author} {\bibinfo {author} {\bibfnamefont {M.}~\bibnamefont
  {Banerji}}, \bibinfo {author} {\bibfnamefont {O.}~\bibnamefont {Lahav}},
  \bibinfo {author} {\bibfnamefont {C.~J.}\ \bibnamefont {Lintott}}, \bibinfo
  {author} {\bibfnamefont {F.~B.}\ \bibnamefont {Abdalla}}, \bibinfo {author}
  {\bibfnamefont {K.}~\bibnamefont {Schawinski}}, \bibinfo {author}
  {\bibfnamefont {S.~P.}\ \bibnamefont {Bamford}}, \bibinfo {author}
  {\bibfnamefont {D.}~\bibnamefont {Andreescu}}, \bibinfo {author}
  {\bibfnamefont {P.}~\bibnamefont {Murray}}, \bibinfo {author} {\bibfnamefont
  {M.~J.}\ \bibnamefont {Raddick}}, \bibinfo {author} {\bibfnamefont
  {A.}~\bibnamefont {Slosar}}, \bibinfo {author} {\bibfnamefont
  {A.}~\bibnamefont {Szalay}}, \bibinfo {author} {\bibfnamefont
  {D.}~\bibnamefont {Thomas}}, \ and\ \bibinfo {author} {\bibfnamefont
  {J.}~\bibnamefont {Vandenberg}},\ }\href {\doibase
  10.1111/j.1365-2966.2010.16713.x} {\bibfield  {journal} {\bibinfo  {journal}
  {Monthly Notices of the Royal Astronomical Society}\ }\textbf {\bibinfo
  {volume} {406}},\ \bibinfo {pages} {342} (\bibinfo {year}
  {2010})}\BibitemShut {NoStop}%
\bibitem [{\citenamefont {Hezaveh}\ \emph {et~al.}(2017)\citenamefont
  {Hezaveh}, \citenamefont {Levasseur},\ and\ \citenamefont
  {Marshall}}]{hezavehFastAutomatedAnalysis2017}%
  \BibitemOpen
  \bibfield  {author} {\bibinfo {author} {\bibfnamefont {Y.~D.}\ \bibnamefont
  {Hezaveh}}, \bibinfo {author} {\bibfnamefont {L.~P.}\ \bibnamefont
  {Levasseur}}, \ and\ \bibinfo {author} {\bibfnamefont {P.~J.}\ \bibnamefont
  {Marshall}},\ }\href {\doibase 10.1038/nature23463} {\bibfield  {journal}
  {\bibinfo  {journal} {Nature}\ }\textbf {\bibinfo {volume} {548}},\ \bibinfo
  {pages} {555} (\bibinfo {year} {2017})}\BibitemShut {NoStop}%
\bibitem [{\citenamefont {Tshitoyan}\ \emph {et~al.}(2019)\citenamefont
  {Tshitoyan}, \citenamefont {Dagdelen}, \citenamefont {Weston}, \citenamefont
  {Dunn}, \citenamefont {Rong}, \citenamefont {Kononova}, \citenamefont
  {Persson}, \citenamefont {Ceder},\ and\ \citenamefont
  {Jain}}]{tshitoyanUnsupervisedWordEmbeddings2019}%
  \BibitemOpen
  \bibfield  {author} {\bibinfo {author} {\bibfnamefont {V.}~\bibnamefont
  {Tshitoyan}}, \bibinfo {author} {\bibfnamefont {J.}~\bibnamefont {Dagdelen}},
  \bibinfo {author} {\bibfnamefont {L.}~\bibnamefont {Weston}}, \bibinfo
  {author} {\bibfnamefont {A.}~\bibnamefont {Dunn}}, \bibinfo {author}
  {\bibfnamefont {Z.}~\bibnamefont {Rong}}, \bibinfo {author} {\bibfnamefont
  {O.}~\bibnamefont {Kononova}}, \bibinfo {author} {\bibfnamefont {K.~A.}\
  \bibnamefont {Persson}}, \bibinfo {author} {\bibfnamefont {G.}~\bibnamefont
  {Ceder}}, \ and\ \bibinfo {author} {\bibfnamefont {A.}~\bibnamefont {Jain}},\
  }\href {\doibase 10.1038/s41586-019-1335-8} {\bibfield  {journal} {\bibinfo
  {journal} {Nature}\ }\textbf {\bibinfo {volume} {571}},\ \bibinfo {pages}
  {95} (\bibinfo {year} {2019})}\BibitemShut {NoStop}%
\bibitem [{\citenamefont {F{\"o}sel}\ \emph {et~al.}(2018)\citenamefont
  {F{\"o}sel}, \citenamefont {Tighineanu}, \citenamefont {Weiss},\ and\
  \citenamefont {Marquardt}}]{foselReinforcementLearningNeural2018}%
  \BibitemOpen
  \bibfield  {author} {\bibinfo {author} {\bibfnamefont {T.}~\bibnamefont
  {F{\"o}sel}}, \bibinfo {author} {\bibfnamefont {P.}~\bibnamefont
  {Tighineanu}}, \bibinfo {author} {\bibfnamefont {T.}~\bibnamefont {Weiss}}, \
  and\ \bibinfo {author} {\bibfnamefont {F.}~\bibnamefont {Marquardt}},\ }\href
  {\doibase 10.1103/PhysRevX.8.031084} {\bibfield  {journal} {\bibinfo
  {journal} {Physical Review X}\ }\textbf {\bibinfo {volume} {8}},\ \bibinfo
  {pages} {031084} (\bibinfo {year} {2018})}\BibitemShut {NoStop}%
\bibitem [{\citenamefont {Kalantre}\ \emph {et~al.}(2019)\citenamefont
  {Kalantre}, \citenamefont {Zwolak}, \citenamefont {Ragole}, \citenamefont
  {Wu}, \citenamefont {Zimmerman}, \citenamefont {Stewart},\ and\ \citenamefont
  {Taylor}}]{kalantreMachineLearningTechniques2019}%
  \BibitemOpen
  \bibfield  {author} {\bibinfo {author} {\bibfnamefont {S.~S.}\ \bibnamefont
  {Kalantre}}, \bibinfo {author} {\bibfnamefont {J.~P.}\ \bibnamefont
  {Zwolak}}, \bibinfo {author} {\bibfnamefont {S.}~\bibnamefont {Ragole}},
  \bibinfo {author} {\bibfnamefont {X.}~\bibnamefont {Wu}}, \bibinfo {author}
  {\bibfnamefont {N.~M.}\ \bibnamefont {Zimmerman}}, \bibinfo {author}
  {\bibfnamefont {M.~D.}\ \bibnamefont {Stewart}}, \ and\ \bibinfo {author}
  {\bibfnamefont {J.~M.}\ \bibnamefont {Taylor}},\ }\href {\doibase
  10.1038/s41534-018-0118-7} {\bibfield  {journal} {\bibinfo  {journal} {npj
  Quantum Information}\ }\textbf {\bibinfo {volume} {5}},\ \bibinfo {pages} {6}
  (\bibinfo {year} {2019})}\BibitemShut {NoStop}%
\bibitem [{\citenamefont {Usman}\ \emph {et~al.}(2019)\citenamefont {Usman},
  \citenamefont {Wong}, \citenamefont {Hill},\ and\ \citenamefont
  {Hollenberg}}]{usmanAtomiclevelCharacterisationQuantum2019}%
  \BibitemOpen
  \bibfield  {author} {\bibinfo {author} {\bibfnamefont {M.}~\bibnamefont
  {Usman}}, \bibinfo {author} {\bibfnamefont {Y.~Z.}\ \bibnamefont {Wong}},
  \bibinfo {author} {\bibfnamefont {C.~D.}\ \bibnamefont {Hill}}, \ and\
  \bibinfo {author} {\bibfnamefont {L.~C.~L.}\ \bibnamefont {Hollenberg}},\
  }\href@noop {} {\bibfield  {journal} {\bibinfo  {journal} {arXiv:1904.01756
  [cond-mat, physics:physics, physics:quant-ph]}\ } (\bibinfo {year} {2019})},\
  \Eprint {http://arxiv.org/abs/1904.01756} {arXiv:1904.01756 [cond-mat,
  physics:physics, physics:quant-ph]} \BibitemShut {NoStop}%
\bibitem [{\citenamefont {Hochreiter}\ and\ \citenamefont
  {Schmidhuber}(1997)}]{hochreiterLongShorttermMemory1997}%
  \BibitemOpen
  \bibfield  {author} {\bibinfo {author} {\bibfnamefont {S.}~\bibnamefont
  {Hochreiter}}\ and\ \bibinfo {author} {\bibfnamefont {J.}~\bibnamefont
  {Schmidhuber}},\ }\href {\doibase 10.1162/neco.1997.9.8.1735} {\bibfield
  {journal} {\bibinfo  {journal} {Neural computation}\ }\textbf {\bibinfo
  {volume} {9}},\ \bibinfo {pages} {1735} (\bibinfo {year} {1997})}\BibitemShut
  {NoStop}%
\bibitem [{\citenamefont {Graves}\ \emph {et~al.}(2013)\citenamefont {Graves},
  \citenamefont {Mohamed},\ and\ \citenamefont
  {Hinton}}]{gravesSpeechRecognitionDeep2013}%
  \BibitemOpen
  \bibfield  {author} {\bibinfo {author} {\bibfnamefont {A.}~\bibnamefont
  {Graves}}, \bibinfo {author} {\bibfnamefont {A.-r.}\ \bibnamefont {Mohamed}},
  \ and\ \bibinfo {author} {\bibfnamefont {G.}~\bibnamefont {Hinton}},\ }in\
  \href {\doibase 10.1109/ICASSP.2013.6638947} {\emph {\bibinfo {booktitle}
  {2013 {{IEEE International Conference}} on {{Acoustics}}, {{Speech}} and
  {{Signal Processing}}}}}\ (\bibinfo {year} {2013})\ pp.\ \bibinfo {pages}
  {6645--6649}\BibitemShut {NoStop}%
\bibitem [{\citenamefont {Graves}\ \emph {et~al.}(2008)\citenamefont {Graves},
  \citenamefont {Liwicki}, \citenamefont {Bunke}, \citenamefont {Schmidhuber},\
  and\ \citenamefont
  {Fern{\'a}ndez}}]{gravesUnconstrainedOnlineHandwriting2008}%
  \BibitemOpen
  \bibfield  {author} {\bibinfo {author} {\bibfnamefont {A.}~\bibnamefont
  {Graves}}, \bibinfo {author} {\bibfnamefont {M.}~\bibnamefont {Liwicki}},
  \bibinfo {author} {\bibfnamefont {H.}~\bibnamefont {Bunke}}, \bibinfo
  {author} {\bibfnamefont {J.}~\bibnamefont {Schmidhuber}}, \ and\ \bibinfo
  {author} {\bibfnamefont {S.}~\bibnamefont {Fern{\'a}ndez}},\ }in\ \href@noop
  {} {\emph {\bibinfo {booktitle} {Advances in {{Neural Information Processing
  Systems}} 20}}},\ \bibinfo {editor} {edited by\ \bibinfo {editor}
  {\bibfnamefont {J.~C.}\ \bibnamefont {Platt}}, \bibinfo {editor}
  {\bibfnamefont {D.}~\bibnamefont {Koller}}, \bibinfo {editor} {\bibfnamefont
  {Y.}~\bibnamefont {Singer}}, \ and\ \bibinfo {editor} {\bibfnamefont {S.~T.}\
  \bibnamefont {Roweis}}}\ (\bibinfo  {publisher} {{Curran Associates, Inc.}},\
  \bibinfo {year} {2008})\ pp.\ \bibinfo {pages} {577--584}\BibitemShut
  {NoStop}%
\bibitem [{\citenamefont {Eck}\ and\ \citenamefont
  {Schmidhuber}(2002)}]{eckLearningLongTermStructure2002}%
  \BibitemOpen
  \bibfield  {author} {\bibinfo {author} {\bibfnamefont {D.}~\bibnamefont
  {Eck}}\ and\ \bibinfo {author} {\bibfnamefont {J.}~\bibnamefont
  {Schmidhuber}},\ }in\ \href {\doibase 10.1007/3-540-46084-5_47} {\emph
  {\bibinfo {booktitle} {Artificial {{Neural Networks}} \textemdash{} {{ICANN}}
  2002}}},\ Vol.\ \bibinfo {volume} {2415},\ \bibinfo {editor} {edited by\
  \bibinfo {editor} {\bibfnamefont {G.}~\bibnamefont {Goos}}, \bibinfo {editor}
  {\bibfnamefont {J.}~\bibnamefont {Hartmanis}}, \bibinfo {editor}
  {\bibfnamefont {J.}~\bibnamefont {{van Leeuwen}}}, \ and\ \bibinfo {editor}
  {\bibfnamefont {J.~R.}\ \bibnamefont {Dorronsoro}}}\ (\bibinfo  {publisher}
  {{Springer Berlin Heidelberg}},\ \bibinfo {address} {{Berlin, Heidelberg}},\
  \bibinfo {year} {2002})\ pp.\ \bibinfo {pages} {284--289}\BibitemShut
  {NoStop}%
\bibitem [{\citenamefont {Clevert}\ \emph {et~al.}(2015)\citenamefont
  {Clevert}, \citenamefont {Unterthiner},\ and\ \citenamefont
  {Hochreiter}}]{clevertFastAccurateDeep2015}%
  \BibitemOpen
  \bibfield  {author} {\bibinfo {author} {\bibfnamefont {D.-A.}\ \bibnamefont
  {Clevert}}, \bibinfo {author} {\bibfnamefont {T.}~\bibnamefont
  {Unterthiner}}, \ and\ \bibinfo {author} {\bibfnamefont {S.}~\bibnamefont
  {Hochreiter}},\ }\href@noop {} {\bibfield  {journal} {\bibinfo  {journal}
  {arXiv:1511.07289 [cs]}\ } (\bibinfo {year} {2015})},\ \Eprint
  {http://arxiv.org/abs/1511.07289} {arXiv:1511.07289 [cs]} \BibitemShut
  {NoStop}%
\bibitem [{\citenamefont {Chollet}\ \emph {et~al.}(2015)\citenamefont {Chollet}
  \emph {et~al.}}]{chollet2015keras}%
  \BibitemOpen
  \bibfield  {author} {\bibinfo {author} {\bibfnamefont {F.}~\bibnamefont
  {Chollet}} \emph {et~al.},\ }\href@noop {} {\bibfield  {journal} {\bibinfo
  {journal} {https://keras.io}\ } (\bibinfo {year} {2015})}\BibitemShut
  {NoStop}%
\bibitem [{\citenamefont {Lambert}\ \emph {et~al.}(2014)\citenamefont
  {Lambert}, \citenamefont {Edwards}, \citenamefont {Esmail}, \citenamefont
  {Pollock}, \citenamefont {Barrett}, \citenamefont {Lovett},\ and\
  \citenamefont {Ferguson}}]{lambertExperimentalObservationBreaking2014}%
  \BibitemOpen
  \bibfield  {author} {\bibinfo {author} {\bibfnamefont {N.~J.}\ \bibnamefont
  {Lambert}}, \bibinfo {author} {\bibfnamefont {M.}~\bibnamefont {Edwards}},
  \bibinfo {author} {\bibfnamefont {A.~A.}\ \bibnamefont {Esmail}}, \bibinfo
  {author} {\bibfnamefont {F.~A.}\ \bibnamefont {Pollock}}, \bibinfo {author}
  {\bibfnamefont {S.~D.}\ \bibnamefont {Barrett}}, \bibinfo {author}
  {\bibfnamefont {B.~W.}\ \bibnamefont {Lovett}}, \ and\ \bibinfo {author}
  {\bibfnamefont {A.~J.}\ \bibnamefont {Ferguson}},\ }\href {\doibase
  10.1103/PhysRevB.90.140503} {\bibfield  {journal} {\bibinfo  {journal}
  {Physical Review B}\ }\textbf {\bibinfo {volume} {90}},\ \bibinfo {pages}
  {140503(R)} (\bibinfo {year} {2014})}\BibitemShut {NoStop}%
\bibitem [{\citenamefont {Lambert}\ \emph
  {et~al.}(2017{\natexlab{b}})\citenamefont {Lambert}, \citenamefont {Esmail},
  \citenamefont {Pollock}, \citenamefont {Edwards}, \citenamefont {Lovett},\
  and\ \citenamefont
  {Ferguson}}]{lambertMicrowaveIrradiationQuasiparticles2017a}%
  \BibitemOpen
  \bibfield  {author} {\bibinfo {author} {\bibfnamefont {N.~J.}\ \bibnamefont
  {Lambert}}, \bibinfo {author} {\bibfnamefont {A.~A.}\ \bibnamefont {Esmail}},
  \bibinfo {author} {\bibfnamefont {F.~A.}\ \bibnamefont {Pollock}}, \bibinfo
  {author} {\bibfnamefont {M.}~\bibnamefont {Edwards}}, \bibinfo {author}
  {\bibfnamefont {B.~W.}\ \bibnamefont {Lovett}}, \ and\ \bibinfo {author}
  {\bibfnamefont {A.~J.}\ \bibnamefont {Ferguson}},\ }\href {\doibase
  10.1103/PhysRevB.95.235413} {\bibfield  {journal} {\bibinfo  {journal}
  {Physical Review B}\ }\textbf {\bibinfo {volume} {95}},\ \bibinfo {pages}
  {235413} (\bibinfo {year} {2017}{\natexlab{b}})}\BibitemShut {NoStop}%
\bibitem [{\citenamefont {Naaman}\ and\ \citenamefont
  {Aumentado}(2006{\natexlab{b}})}]{naamanTimedomainMeasurementsQuasiparticle2006}%
  \BibitemOpen
  \bibfield  {author} {\bibinfo {author} {\bibfnamefont {O.}~\bibnamefont
  {Naaman}}\ and\ \bibinfo {author} {\bibfnamefont {J.}~\bibnamefont
  {Aumentado}},\ }\href {\doibase 10.1103/PhysRevB.73.172504} {\bibfield
  {journal} {\bibinfo  {journal} {Physical Review B}\ }\textbf {\bibinfo
  {volume} {73}},\ \bibinfo {pages} {172504} (\bibinfo {year}
  {2006}{\natexlab{b}})}\BibitemShut {NoStop}%
\bibitem [{\citenamefont {Ferguson}\ \emph {et~al.}(2006)\citenamefont
  {Ferguson}, \citenamefont {Court}, \citenamefont {Hudson},\ and\
  \citenamefont {Clark}}]{fergusonMicrosecondResolutionQuasiparticle2006}%
  \BibitemOpen
  \bibfield  {author} {\bibinfo {author} {\bibfnamefont {A.~J.}\ \bibnamefont
  {Ferguson}}, \bibinfo {author} {\bibfnamefont {N.~A.}\ \bibnamefont {Court}},
  \bibinfo {author} {\bibfnamefont {F.~E.}\ \bibnamefont {Hudson}}, \ and\
  \bibinfo {author} {\bibfnamefont {R.~G.}\ \bibnamefont {Clark}},\ }\href
  {\doibase 10.1103/PhysRevLett.97.106603} {\bibfield  {journal} {\bibinfo
  {journal} {Physical Review Letters}\ }\textbf {\bibinfo {volume} {97}},\
  \bibinfo {pages} {106603} (\bibinfo {year} {2006})}\BibitemShut {NoStop}%
\bibitem [{\citenamefont {Lutchyn}\ \emph {et~al.}(2005)\citenamefont
  {Lutchyn}, \citenamefont {Glazman},\ and\ \citenamefont
  {Larkin}}]{lutchynQuasiparticleDecayRate2005}%
  \BibitemOpen
  \bibfield  {author} {\bibinfo {author} {\bibfnamefont {R.}~\bibnamefont
  {Lutchyn}}, \bibinfo {author} {\bibfnamefont {L.}~\bibnamefont {Glazman}}, \
  and\ \bibinfo {author} {\bibfnamefont {A.}~\bibnamefont {Larkin}},\ }\href
  {\doibase 10.1103/PhysRevB.72.014517} {\bibfield  {journal} {\bibinfo
  {journal} {Physical Review B}\ }\textbf {\bibinfo {volume} {72}},\ \bibinfo
  {pages} {014517} (\bibinfo {year} {2005})}\BibitemShut {NoStop}%
\bibitem [{\citenamefont {Esmail}\ \emph {et~al.}(2017)\citenamefont {Esmail},
  \citenamefont {Ferguson},\ and\ \citenamefont
  {Lambert}}]{esmailCooperPairTunnelling2017}%
  \BibitemOpen
  \bibfield  {author} {\bibinfo {author} {\bibfnamefont {A.~A.}\ \bibnamefont
  {Esmail}}, \bibinfo {author} {\bibfnamefont {A.~J.}\ \bibnamefont
  {Ferguson}}, \ and\ \bibinfo {author} {\bibfnamefont {N.~J.}\ \bibnamefont
  {Lambert}},\ }\href {\doibase 10.1063/1.5009079} {\bibfield  {journal}
  {\bibinfo  {journal} {Applied Physics Letters}\ }\textbf {\bibinfo {volume}
  {111}},\ \bibinfo {pages} {252602} (\bibinfo {year} {2017})}\BibitemShut
  {NoStop}%
\end{thebibliography}%

\end{document}